# Enhanced triplet superconductivity in next-generation ultraclean UTe$_2$

Z. Wu,[1][†] T. I. Weinberger,[1][†] J. Chen,[1] A. Cabala,[2] D. V. Chichinadze,[3]
D. Shaffer,[4] J. Pospíšil,[2] J. Prokleška,[2] T. Haidamak,[2] G. Bastien,[2]
V. Sechovský,[2] A. J. Hickey,[1] M. J. Mancera-Ugarte,[5] S. Benjamin,[3]
D. E. Graf,[3] Y. Skourski,[6] G. G. Lonzarich,[1] M. Vališka,[2]
F. M. Grosche,[1] A. G. Eaton[1][*].

[1]Cavendish Laboratory, University of Cambridge,
JJ Thomson Avenue, Cambridge, CB3 0HE, UK
[2]Charles University, Faculty of Mathematics and Physics, Department of
Condensed Matter Physics, Ke Karlovu 5, Prague 2, 121 16, Czech Republic
[3]National High Magnetic Field Laboratory, Tallahassee, Florida 32310, USA
[4]Department of Physics, Emory University, 400 Dowman Drive, Atlanta, Georgia 30322, USA
[5]Department of Chemical Engineering and Biotechnology,
University of Cambridge, Cambridge, CB3 0AS, United Kingdom
[6]Hochfeld-Magnetlabor Dresden (HLD-EMFL),
Helmholtz-Zentrum Dresden-Rossendorf, Dresden, 01328, Germany

[*]Email: alex.eaton@phy.cam.ac.uk.
[†]These authors contributed equally to this work.

**The unconventional superconductor UTe$_2$ exhibits numerous signatures of spin-triplet superconductivity – a rare state of matter which could enable quantum computation protected against decoherence. UTe$_2$ possesses a complex phase landscape comprising two magnetic field-induced superconducting phases, a metamagnetic transition to a field-polarised state, along with pair- and charge-density wave orders. However, contradictory reports between studies performed on UTe$_2$ specimens of varying quality have severely impeded theoretical efforts to understand the microscopic origins of the exotic superconductivity. Here, we report a comprehensive suite of high magnetic field measurements on a new generation of pristine quality UTe$_2$ crystals. Our experiments reveal a significantly revised high magnetic field superconducting phase diagram in the ultraclean limit, showing a pronounced sensitivity of field-induced superconductivity to the presence of crystalline disorder. We employ a Ginzburg-Landau model that excellently captures this acute dependence on sample quality. Our results suggest that in close proximity to a field–induced metamagnetic transition the enhanced role of magnetic fluctuations – that are strongly suppressed by disorder – is likely responsible for tuning UTe$_2$ between two distinct spin-triplet superconducting phases.**

# Introduction

A superconducting state is attained when a material exhibits macroscopic quantum phase coherence. Conventional (BCS) superconductors possess a bosonic coherent quantum fluid composed of pairs of electrons that are weakly bound together by phononic mediation to form a Cooper pair.[1,2] The condensation of Cooper pairs also drives superconductivity in unconventional superconductors, but in these materials the pairing glue originates not from phonons but instead from attractive interactions typically found on the border of density or magnetic instabilities.[3] The majority of known unconventional superconductors exhibit magnetically mediated superconductivity located in close proximity to an antiferromagnetically ordered state, comprising Cooper pairs in a spin-singlet configuration that have a total charge of 2$e$ and zero net spin.[4,5]

The discovery of superconductivity in the ferromagnets UGe$_2$,[6] URhGe,[7] and UCoGe[8] was surprising because most superconducting states are fragile to the presence of a magnetic field, as this tends to break apart the Cooper pairs that compose the charged superfluid. However, an alternative pairing mechanism was proposed for these materials, involving two electrons of the same spin combined in a triplet configuration, for which ferromagnetic correlations may thus enhance an attractive pair-forming interaction.[9]

The discovery of superconductivity below 1.6 K in UTe$_2$[10] was also met with surprise, as although this material also exhibits several features characteristic of spin-triplet pairing, it possesses a paramagnetic rather than ferromagnetic groundstate. Two of the strongest observations in favor of triplet superconductivity in UTe$_2$ include a small change in the NMR Knight shift on cooling through the superconducting critical temperature ($T_c$), and large upper critical fields along each crystallographic axis that are considerably higher than the Pauli-limit for spin-singlet Cooper pairs.[11] Notably, for a magnetic field, $H$, applied along the hard magnetic $b$ direction, superconductivity persists to $\mu_0 H \approx 35$ T – over an order of magnitude higher than the Pauli limit,[12,13] at which point it is sharply truncated by a first-order metamagnetic (MM) transition into a field-polarised phase.[14,15] Remarkably, this field-polarised state hosts a magnetic field-reentrant superconducting phase over a narrow angular range of applied field, which onsets at $\mu_0 H \approx 40$ T[14,16,17] and appears to persist to $\mu_0 H \approx 70$ T.[18]

Careful angle-dependent resistivity measurements in high magnetic fields, for field applied in close proximity to the $b$-axis, observed the appearance of two distinct superconducting phases over the field interval of $0$ T $\leq \mu_0 H \lessapprox 35$ T.[14,15] This interpretation has recently been corroborated by bulk thermodynamic measurements at this field orientation, indicating the presence of a distinct field-reinforced superconducting state for $\mu_0 H \gtrapprox 15$ T.[19] Throughout this report we shall refer to the zero field superconducting state as SC1, to the field-reinforced phase for field applied close to the $b$ direction as SC2, and to the very high magnetic field-induced phase, located at $\mu_0 H \gtrapprox 40$ T for inclined angles in the $b - c$ rotation plane, as SC3.

Several early studies of the superconducting properties of UTe$_2$ observed two superconducting transitions in the temperature dependence of the specific heat (in zero applied magnetic field),[10,20,21] leading to speculation regarding a possible multi-component nature of the superconducting order parameter at ambient pressure and magnetic field. However, subsequent reports demonstrated that this was perhaps instead an artifact of sample inhomogeneity,[11,22] with higher quality samples found to exhibit a singular sharp superconducting transition.[23–25] Kerr effect measurements on samples exhibiting two specific heat transitions yielded evidence for time reversal symmetry breaking;[20] however, this observation could not be reproduced on higher quality samples.[26] Theoretical efforts to understand the microscopic details of the remarkable superconducting properties of UTe$_2$ have thus been stymied by these discrepancies between experimental studies performed on samples of varying quality.

In addition to the three superconducting phases and the high field spin-polarised state, UTe$_2$ also possesses pair density wave (PDW)[27] and charge density wave (CDW)[28] ordering. Unusu-ally, the CDW state appears to be fragile to the application of a magnetic field, and has been reported to terminate at the upper critical field ($H_{c2}$) of the SC1 state.[28] Given this rich variety of exotic electronic phases, a more detailed understanding of the phase landscape – in high quality samples – is urgently called for in order to guide theoretical efforts in their attempt to better understand the interplay between this assortment of strongly correlated electronic states.

In this work we report measurements on a new generation of UTe$_2$ crystals grown by a molten salt flux (MSF) technique, using starting materials of elemental uranium refined by the solid state electrotransport technique[29] and tellurium pieces of 6N purity. The pristine quality of the resulting single crystals is evidenced by their high $T_c$ values of up to 2.10 K, low residual resistivities down to 0.48 $\mu\Omega$ cm, and the observation of magnetic quantum oscillations at high magnetic fields and low temperatures.[25] Concomitant with the enhancement in $T_c$, the $H_{c2}$ values of SC1 along the *a* and *c* directions are also enhanced in comparison to samples with lower $T_c$ values. Remarkably, we find that the angular extent of SC2 – that is, the rotation angle away from *b* over which a zero resistance state is still observed at low temperatures for $\mu_0 H \approx 30$ T – is significantly enhanced for this new generation of high purity crystals. We propose a model to capture this behaviour, and find that our observations can be well described by considering the enhanced role of magnetic fluctuations close to the MM transition.

By contrast, we find that the MM transition to the field polarised state still sharply truncates superconductivity at $\mu_0 H_m \approx 35$ T in our high quality MSF samples. This indicates that while the SC1 and SC2 superconducting phases of UTe$_2$ are highly sensitive to the effects of crystalline disorder, the first-order phase transition to the high magnetic field polarised paramagnetic state is an intrinsic magnetic feature of the UTe$_2$ system, and is robust against disorder. We also find that the formation of the SC3 phase in ultraclean MSF samples appears to follow the same field-angle profile found in prior sample generations grown by the chemical vapor transport (CVT) method.

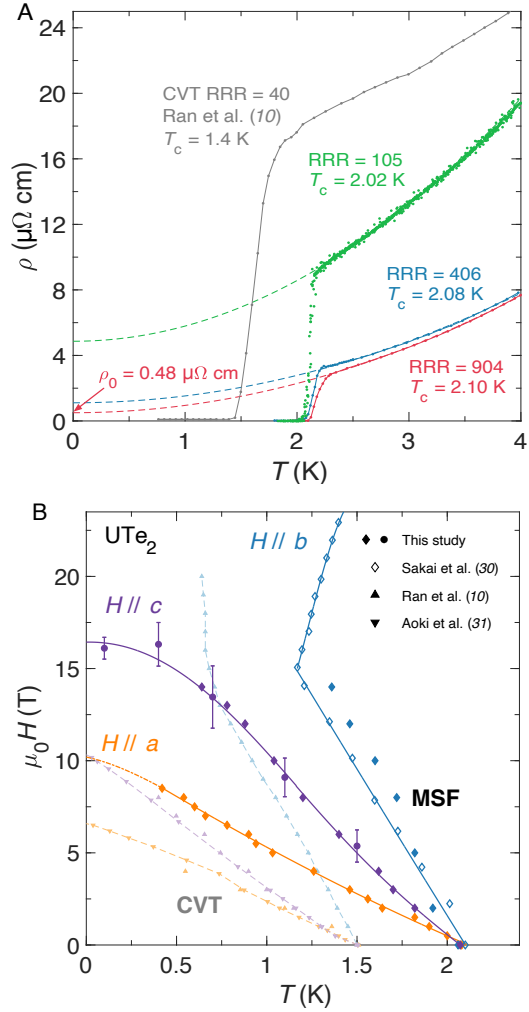

**Fig. 1.** Sensitivity of superconductivity to crystalline disorder. (*A*) Electrical resistivity, $\rho$, as a function of temperature, $T$, for three samples grown by the molten salt flux (MSF) technique (coloured points), plotted alongside data reported for a chemical vapor transport (CVT) specimen in ref.[10] $T_c$ values were determined by zero resistivity, as defined in Table 1. Residual resistivity ratios (RRRs) were computed by fitting the low temperature normal state resistivity with the dashed curves, of functional form $\rho = AT^2 + \rho_0$ for constant $A$, to extract the residual normal state resistivity $\rho_0$. The dimensionless RRR value is defined as $\rho(T = 300\text{ K})/\rho_0$. (*B*) Magnetic field–temperature superconducting phase diagram of UTe$_2$. For field oriented along each crystallographic axis, $T_c(H)$ is enhanced for MSF samples (bold symbols) in comparison to CVT samples (pale symbols). Lines are given as a guide to the eye. Contacted (contactless) resistivity measurements from this study are represented by solid diamonds (circles). Raw resistivity data used in part to construct this figure are given in the Supporting Information, as is the procedure for determining error bars for contactless resistivity points. All contacted resistivity measurements were performed on the RRR = 406 sample from Table 1. Additional MSF resistivity data along the $b$ direction are reproduced from ref.[30] CVT resistivity data are given by up (down) triangles, reproduced from ref.[10] (ref.[31]). We identify the normal-superconducting transition temperature by the point at which zero resistivity is first attained (as defined in Table 1).

# Results

## Enhancement of $T_c$ and $H_{c2}$ of SC1

Figure 1 shows the temperature dependence of the electrical resistivity, $\rho(T)$, for three MSF samples (coloured points) of varying quality. Data for $\rho(T)$ of a CVT sample reported in ref.[10] is plotted in gray for comparison. A clear trend is apparent, with samples exhibiting higher $T_c$ values also possessing higher residual resistivity ratios (RRRs), where the RRR is the ratio between the residual resistivity, $\rho_0$, and $\rho(T = 300\text{ K})$.

Table 1 tabulates these data presented in Fig. 1a, and also includes data from other studies as indicated. Here, the correlation between $T_c$ and RRR is further emphasised, with samples exhibiting high $T_c$ values also possessing low residual resistivities (and thus high RRRs). A high RRR is indicative of high sample purity,[23] as samples containing less crystalline disorder will thus have lower scattering rates for the itinerant quasiparticles partaking in the electrical transport measurement. Characterising sample quality by comparison of RRR values is a particularly effective methodology, as it is agnostic with regards to the source of the crystalline disorder – be it from grain boundaries or vacancies or impurities, from some other source of disorder, or indeed a combination of several types. The presence of any such defects will lead to an increase in the charge carrier scattering rate, thereby yielding a lower resultant RRR.

Fig. 1b shows a comparison of the extent of superconductivity for CVT and MSF samples. For magnetic field applied along the crystallographic $a$ and $c$ directions, $H_{c2}$ is clearly enhanced for the cleaner MSF samples, in good agreement with ref.[33] Along the hard magnetic $b$ direction, $T_c(H)$ is also enhanced for all temperatures measured. The effect of magnetic field-reinforced superconductivity along this direction is observed as a kink in the $T_c(H)$ curve at $\mu_0 H \approx 15$ T, as reported previously[14,19] – but this feature occurs at higher temperature in the case of MSF-grown UTe$_2$ compared to CVT samples. We also find that the lower criti-cal field ($H_{c1}$) is enhanced for MSF samples, consistent with a recent report[34] (see Supporting Information).

**Table 1:** Comparison of critical superconducting temperature ($T_c$), residual resistivity ($\rho_0$), and the residual resistivity ratio (RRR) for UTe$_2$ samples grown by the MSF and CVT techniques from various reports as indicated. In all cases, $T_c$ is defined by zero resistivity, which we identify as the first measurement point to fall below 0.1 μΩ cm on cooling. $\rho_0$ is determined by a quadratic fitting at low temperatures, as depicted in Figure 1, to give the expected normal state resistivity value at 0 K in the absence of superconductivity. RRR is the ratio between $\rho_0$ and $\rho(T = 300 \text{ K})$. FIB stands for focused ion beam. Note that in Sakai et al.[32] the authors stated that their RRR = 1000 sample was too small to accurately determine the resistivity – therefore a value for $\rho_0$ was not obtained.

| Growth method | $T_c$ (K) | $\rho_0$ (μΩ cm) | RRR | Reference |
|---|---|---|---|---|
| MSF | 2.10 | 0.48 | 904 | This study |
|  | 2.08 | 1.1 | 406 |  |
|  | 2.02 | 4.7 | 105 |  |
| MSF | 2.06 | 1.7 | 220 | Aoki et al. (2022)[24] |
| MSF | 2.10 | - | 1000 | Sakai et al. (2022)[32] |
|  | 2.04 | 2.4 | 170 |  |
| CVT | 2.00 | 7 | 88 | Rosa et al. (2022)[23] |
|  | 1.95 | 9 | 70 |  |
|  | 1.85 | 12 | 55 |  |
| CVT | 1.44 | 16 | 40 | Ran et al. (2019)[10] |
| CVT | 1.55 - 1.60 | 19 | 35 | Aoki et al. (2019)[31] |
| CVT | 1.55 - 1.60 | 16 | 35 - 40 | Helm et al. (2022)[18] |
| CVT FIB | 1.55 - 1.60 | 27 | 25 - 30 |  |

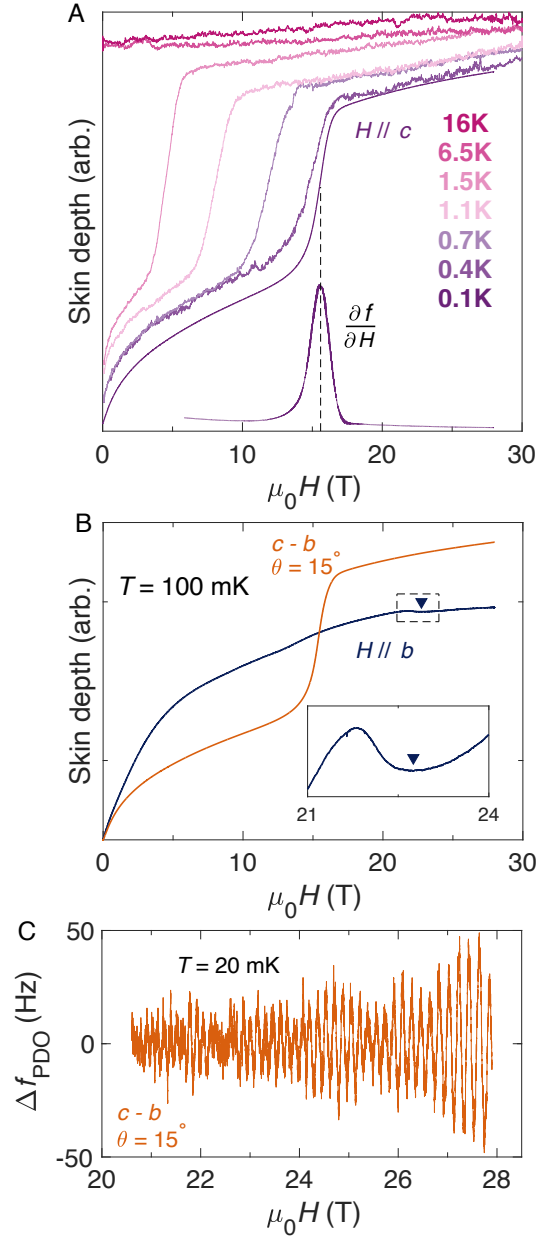

**Fig. 2.** Skin depth measurements of pristine UTe$_2$. (*A*) PDO measurement (see Methods) of the skin depth of UTe$_2$ for magnetic field applied along the $c$ direction at various temperatures (strictly, this is a measurement of $\frac{\Delta f}{f}$ as per Eq. 4, which we refer to as skin depth for succinctness). The derivative of the 0.1 K curve is also plotted ($\frac{\partial f_{PDO}}{\partial H}$), identifying the superconducting transition out of the SC1 state. These data form part of Fig 1. (*B*) Skin depth for field oriented along the $b$ direction (dark blue curve) and tilted 15° from $c$ towards $b$ (ochre curve). The inset shows a zoomed view of the $H \parallel b$ data, with an arrow marking the location of an anomalous feature that appears to indicate the boundary between SC1 and SC2. (*C*) Oscillatory component of the PDO signal at 20 mK, showing prominent quantum oscillations of frequencies $\approx$ 3.5 kT, consistent with prior studies.[24,25] Data at $T \geq 0.4$ K were measured in a resistive magnet, with the lower temperature measurements performed in a superconducting magnet. All data in this figure were collected on the same sample.

This observation of increased sample purity leading to an enhancement of $T_c$ and $H_c$ is not uncommon for unconventional superconductors, with a strong correlation between $T_c$ and $\rho_0$ previously reported, for example, in studies of ruthenates,[35] cuprates,[36] and heavy fermion superconductors.[37,38] A quantitative analysis of the effect of crystalline disorder can often be achieved by utilizing the Abrikosov-Gor'kov theory.[39] However, it has been suggested that this approach may not be valid for the case of UTe$_2$,[40] indicating a complex dependence of superconductivity on the presence of disorder, as may be expected for a $p$-wave superconductor.

The high purity of UTe$_2$ samples investigated in this study is further underlined by their ability to exhibit magnetic quantum oscillations through the de Haas-van Alphen (dHvA) and quantum interference effects at high magnetic fields and low temperatures. All measurements reported in this study were performed on crystals from the same batch as those previously reported[25,41] to exhibit high frequency quantum oscillations, indicative of a long mean free path and thus high crystalline quality.

Figure 2 shows the PDO response (see Methods) of UTe$_2$ at low temperatures up to intermediate magnetic field strengths. Note that the response of the PDO circuit is expressed in full in Eq. 4 – for brevity, we shall refer to this throughout as the skin depth, as aspects of both $\rho$ and $\chi_s$ are important. Fig. 2a maps the superconducting phase boundary for $H \parallel c$. In Fig. 2c the oscillatory component ($\Delta f_{\text{PDO}}$) of the PDO signal at $T = 20$ mK is isolated, which exhibits clear quantum oscillations. The observation of quantum oscillations in a material requires $\omega_c \tau \gtrsim 1$, where $\omega_c$ is the cyclotron frequency and $\tau$ is the quasiparticle lifetime.[42] Therefore, the manifestation of quantum oscillations in our samples indicates that the mapping of the UTe$_2$ phase diagram presented in this study gives an accurate description of the UTe$_2$ system in the clean quantum limit.

## Pronounced angular enhancement of SC2

One of the most remarkable features of the UTe$_2$ phase diagram (at ambient pressure) is the presence of three distinct superconducting phases for magnetic field aligned along certain orientations.[14,43] For $H$ applied along the $b$ direction, at low temperatures ($T < 0.5$ K) zero resistance is observed all the way up to 34.5 T.[16] Remarkably, at higher temperatures ($T \approx 1$ K) and for field applied at a slight tilt angle away from $H \parallel b$, measurements of CVT samples have shown that rather than a single superconducting state persisting for 0 T $\leq \mu_0 H \leq$ 34.5 T, there are instead two distinct superconducting phases present over this field interval,[19] with the higher-field phase (SC2) having been referred to as a "field-reinforced" superconducting state.[11]

Figure 3 shows the skin depth of UTe$_2$ measured in pulsed magnetic fields up to 70 T, for field applied along the hard magnetic $b$ direction. The MM transition to the polarised paramagnetic state is clearly observed by a sharp step in the skin depth at $\mu_0 H_m \approx 35$ T for all temperatures.[11] An interesting aspect of our PDO measurements is the presence of an anomalous kink feature, marked with arrows in Fig. 3a (and in the inset of Fig. 2b), which appears to demarcate the phase boundary between SC1 and either SC2 or the normal state, depending on the temperature. These points are plotted as purple circles in Fig. 3, along with resistivity and specific heat data from previous reports.[10,16,19,30] By Eq. 4 the change in frequency of the PDO circuit is sensitive to both the electrical resistivity and the magnetic susceptibility of the sample. Thus, this observation appears consistent with recent reports[17,30] in which a kink in the magnetic susceptibility has been attributed to marking the termination of SC1, which is visible in our skin depth measurements even though the resistivity remains zero as the material passes from SC1 to SC2.

Figure 4 shows the resistivity of MSF-grown UTe$_2$ measured in a resistive magnet over the field interval 0 T $\leq \mu_0 H \leq$ 41.5 T at $T = 0.4$ K for various magnetic field tilt angles as indicated. Data in the $b - c$ plane were taken on the RRR = 406 sample from Table 1 while those in the $b-a$ plane are from the RRR = 105 sample. At $T = 0.4$ K, for small tilt angles within 5° from the $b$ direction in both rotation planes, zero resistivity persists until the magnetic field strength exceeds 34.0 T, whereupon the resistivity increases rapidly at the MM transition as SC2 terminates and the polarised paramagnetic state is entered. In the $b - c$ rotation plane, this remains the case for angles up to 19° away from $b$; however, by 25° nonzero resistivity is observed for $\mu_0 H$ as low as 20 T (Fig. 4a). Above 20 T the resistivity at this angle then remains small but nonzero up to 38 T. At this point the SC3 phase is accessed and zero resistivity is observed up to this measurement's highest applied field strength of 41.5 T.

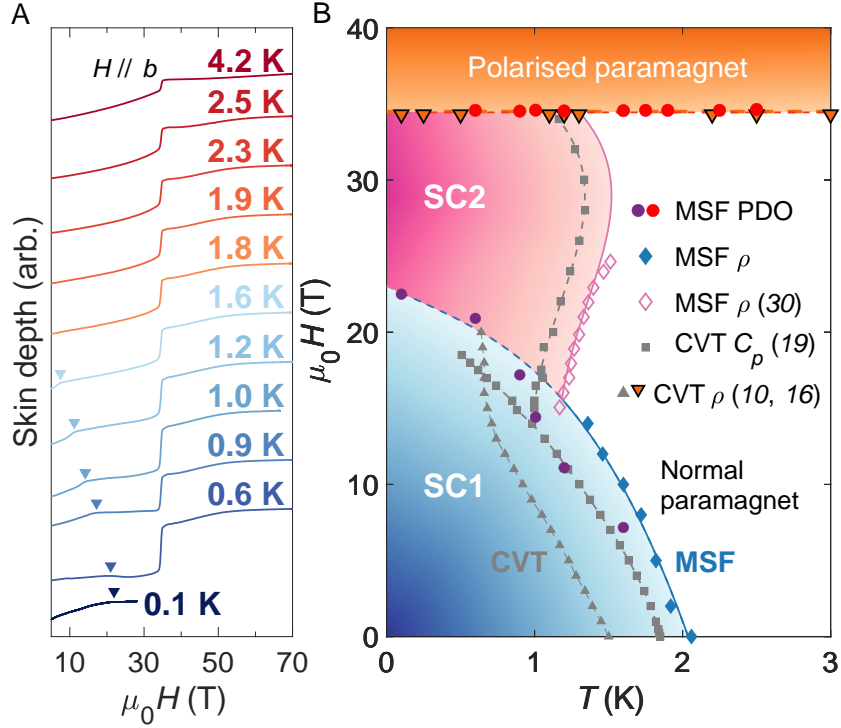

**Fig. 3. Mapping the interplay between SC1, SC2 and metamagnetism for $H \parallel b$.**
(*A*) PDO measurements for $H \parallel b$ at indicated temperatures. The 0.1 K curve is the same data as in Fig. 2b, measured in a dc magnet; all other data were obtained in a pulsed magnet. Arrows indicate the anomalous feature in the PDO signal displayed in Fig. 2b, marked by purple circles in panel (B), which indicates a magnetic field-induced transition between two superconducting states (SC1 and SC2). (*B*) Field-temperature phase diagram comparing the phase-space of CVT and MSF UTe$_2$ samples for $H \parallel b$. Points are from refs[10,16,19,30] as indicated. Lines are as a guide to the eye. Two distinct superconducting phases are observed at low temperatures for this field orientation, which we label as SC1 and SC2. The extent of both SC1 and SC2 in temperature is clearly enhanced for MSF samples compared to CVT specimens. However, both types of samples see the SC2 phase sharply truncated by a MM transition to a field polarised state at $\mu_0 H_m \approx 35$ T.

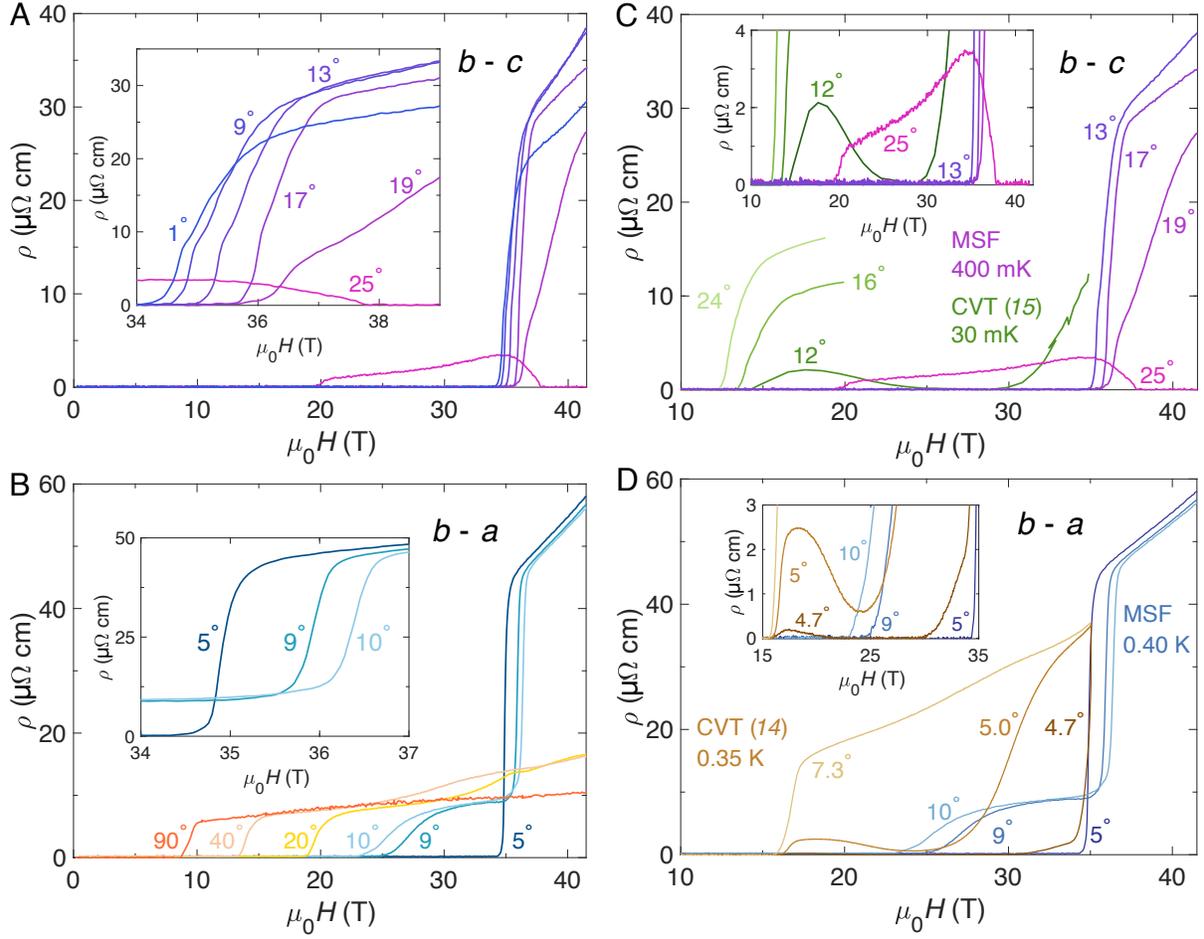

**Fig. 4.** Purity-induced angular enhancement of the SC2 phase. Angular dependence of resistivity for rotation in *(A)* the $b-c$ plane and *(B)* the $b-a$ plane. 0° corresponds to $H \parallel b$ for both panels. Insets give a zoomed view of the magnetic field interval over which the MM transition is located. The data in panel (A) were recorded on the RRR = 406 sample from Table 1 while those in panel (B) are from the RRR = 105 sample. All data were obtained at $T = 0.4$ K. Comparisons of UTe$_2$ $\rho(H)$ data for MSF and CVT samples are given for *(C)* the $b-c$ rotation plane and *(D)* the $b-a$ rotation plane. Insets give a zoomed view of the main panels. MSF curves for selected angles are reproduced from panels (A,B). CVT data in (C) are reproduced from ref.[15] while those in (D) are from ref.[14]

In Fig. 4c,d we compare the angular extent of SC2 by collating selected angles from panels *A,B* alongside prior CVT studies. In the $b - c$ rotation plane, CVT measurements reported by Knebel et al.[15] found that for a rotation angle of 8° away from $b$, zero resistivity persisted up to their highest accessed field strength of 35 T. However, at 12° this was no longer the case, with nonzero resistance observed over the field interval of 14 T $\lessapprox \mu_0 H \lessapprox$ 25 T. The resistivity then returned to zero for 25 T $\lessapprox \mu_0 H \approx$ 30 T, above which it increased up until 35 T (Fig. 4c).

By contrast, our measurements on MSF-grown UTe$_2$ yield zero resistivity over the entire field interval 0 T $\leq \mu_0 H \lessapprox$ 34.5 T for successive tilt angles up to and including 19° away from $b$ towards $c$. Notably, our measurements in the $b - c$ plane were performed in a $^3$He system, at a temperature an order of magnitude higher than those reported by Knebel et al. from dilution fridge measurements.[15] This indicates a remarkable angular expansion of SC2 resulting from the enhancement of purity in this new generation of crystals.

A similar trend is found in the $b - a$ rotation plane. Prior measurements on a CVT specimen reported by Ran et al.[14] found a strong sensitivity of the extent of SC2 within a very small angular range of only 0.3°, with markedly different $\rho(H)$ observed for 4.7° compared to 5.0° (Fig. 4d). By comparison, at 5° we observed zero resistance persisting to $\mu_0 H >$ 34 T, while at 9° and 10° the resistive transition is notably sensitive to small changes in angle, indicating that the boundary of SC2 for MSF samples lies close to here. Interestingly, it appears that the angular extent of SC2 in both rotation planes appears to be approximately doubled for MSF compared to CVT samples – for angles $b-c$ from approximately 12° to between 19°-25°, and for $b-a$ from 5° to around 10°.

## Field-angle phase space of UTe$_2$

The previous sections have demonstrated that the critical fields of SC1, and the angular extent of SC2, have been enhanced for this new generation of pristine quality UTe$_2$ crystals. We turn our attention now to consider the behaviour of the field polarised state, which is instructive as it is this phase into which SC2 is abruptly quenched, and out of which SC3 emerges.

Fig. 3 shows a clear step in the skin depth for $H \parallel b$ at $\mu_0 H \approx 35$ T. Extensive prior high magnetic field measurements on CVT-grown samples have identified this feature as a first-order MM transition to a polarised paramagnetic state at which the magnetization of the material abruptly jumps by $\approx 0.5$ $\mu_B$ per formula unit.[11, 14, 44, 45]

Figure 5 tracks the MM transition as the orientation of the magnetic field is rotated away from $b$ towards $c$, and compares with prior PDO measurements on a CVT specimen reported in ref.[14] At $\theta = \{0°, 20°\}$ the sharp rise in the skin depth – caused by the abrupt increase in resistivity characteristic of entering the polarised paramagnetic phase[43] – occurs at the same value of $H$ for both CVT and MSF samples (within experimental resolution). At $\theta = 33°$, again both samples see a jump in the skin depth at the same field strength – but here the jump is in the opposite direction, due to the presence of SC3.

Figure 6 depicts the phase space of UTe$_2$ for applied magnetic fields oriented in the $b-c$ and $b-a$ planes, at strengths up to 70 T, combining our MSF data with prior CVT studies. CVT $\rho$ from Knebel et al.[15] was reportedly measured at $T = 30$ mK; our MSF PDO points tracking the termination of SC1 were measured at $T = 0.1$ K. All our $\rho$ points in this figure were measured at $T = 0.4$ K in steady fields, while the $\rho$ and PDO measurements reported by Ran et al.[14] were performed both in steady and pulsed fields, at $T \approx 0.4$-$0.5$ K. Our pulsed field

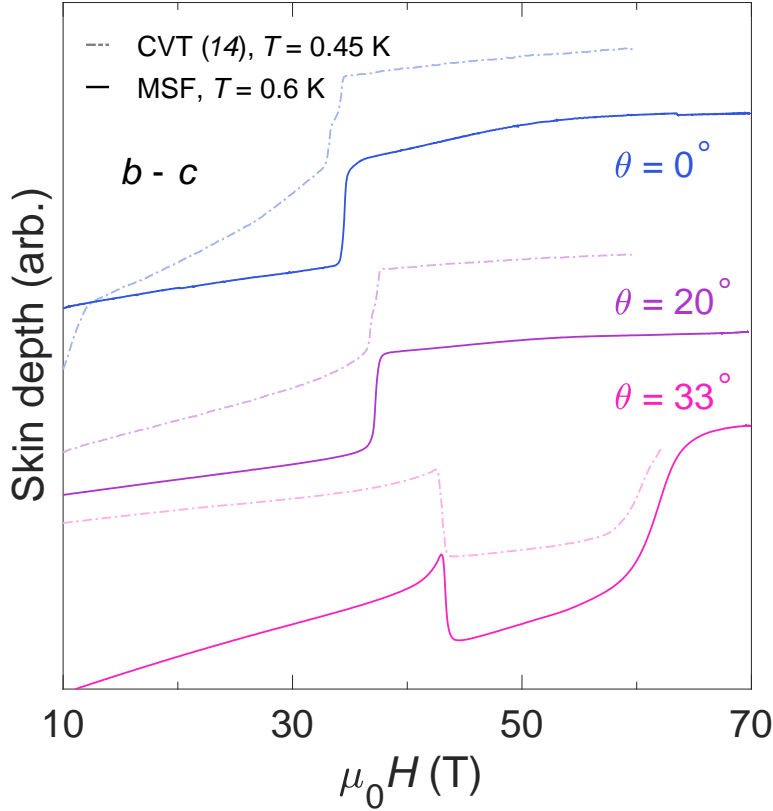

**Fig. 5. Consistency between metamagnetism and SC3 of CVT and MSF UTe$_2$.** Angular evolution of the MM transition at high fields in the $b - c$ plane; $\theta = 0°$ corresponds to $H \parallel b$. Notably, we find that the location of the MM transition is unchanged comparing between MSF (solid curves) and CVT (dashed curves from ref.[14]) samples, including for the onset of re-entrant superconductivity (SC3) at $\theta = 33°$.

PDO measurements tracking the field polarised state, and the $\rho$ measurements reported in Helm et al.,[18] were performed at $T \approx 0.6\text{-}0.7$ K.

Upon inspecting Figs. 5 and 6, there appears to be negligible difference between measurements of the MM transition for MSF and CVT samples. This indicates that this transition is an intrinsic property of the UTe$_2$ system that, unlike SC1 and SC2, is insensitive to crystalline disorder. Furthermore, we find that the temperature evolution of the MM transition tracks very similarly between MSF and CVT samples, implying that the associated energy scale is un-changed under the improvement of sample quality (see Supplementary Figure S7 for steady field data up to $T = 34$ K).[16,46]

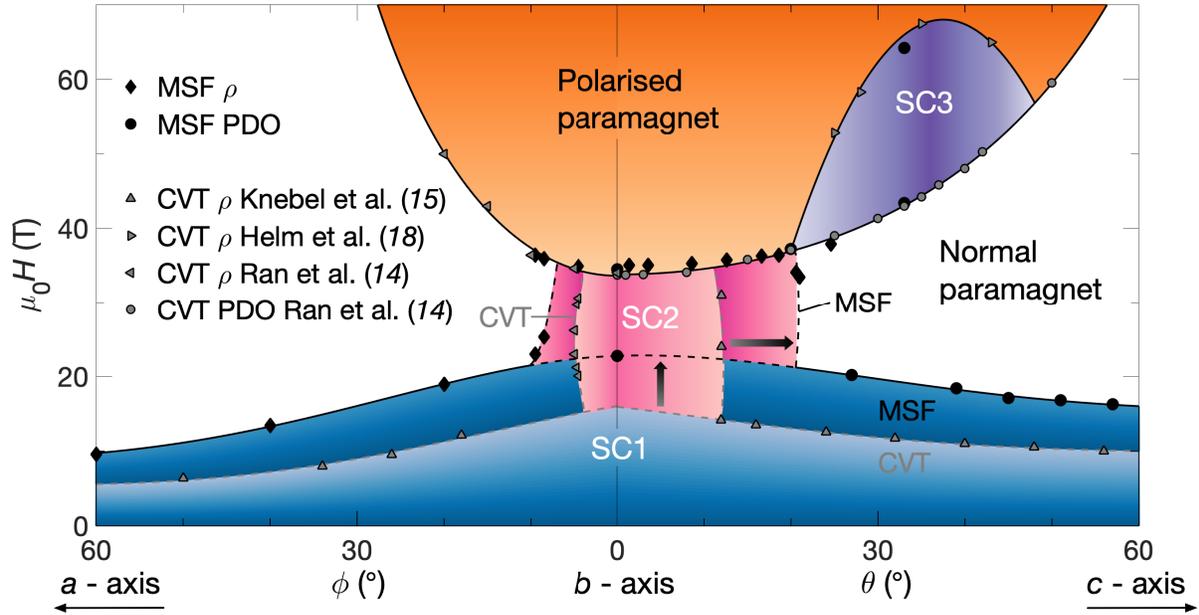

**Fig. 6.** High magnetic field phase diagram for pristine quality MSF-grown $UTe_2$. We find that the phase boundary between SC1 and the normal state is located at higher magnetic field strengths for MSF samples compared to prior studies on CVT specimens (blue region). Furthermore, the angular extent of SC2 is greatly enhanced for MSF samples (pink region). The polarised paramagnetic state (orange region) is found to have the same angular profile for both types of samples. Lines and shading are as a guide to the eye. CVT data points from refs.[14,15,18]

## Modelling the origin of SC2

The mechanism behind, and the precise form of, the superconducting order parameter in $UTe_2$ remains the subject of much theoretical debate.[47–54] The current consensus appears to be that at zero external field a triplet order parameter is stabilized by some form of magnetic fluctuations, giving rise to the SC1 phase.[11] These fluctuations were initially presumed to be ferromagnetic in character[10] but have subsequently been proposed to be antiferromagnetic[55] – for our modelling of the SC1 and SC2 phases we shall remain agnostic as to the precise nature of these low field fluctuations. In sharp contrast to SC1, the experimental data strongly indicate that the SC2 phase has a rather different character, as evidenced by its acute sensitivity to the field direction, its starkly different NMR spectra, and by the observation of $T_c$ growing with increasing field aligned along the $b$-axis.[19,30,54,56–58] Furthermore, the presence of spatially uniform ($Q = 0$) metamagnetic fluctuations has recently been reported[59] for sufficiently strong magnetic fields applied along the $b$-axis.

These observations suggest that the SC2 phase likely has a very different pairing mechanism compared to SC1, with a distinct possibility being that it is driven by the observed MM fluctuations as the first-order transition to the polarised paramagnetic state is approached in high magnetic fields. Such a mechanism for magnetic field-reinforced superconductivity has previously been considered in the case of the ferromagnetic superconductors URhGe and UCoGe.[9,60–62] The presence of strong fluctuations at a strong first order phase transition is unusual, but likely originates from low coercivity of the transition as seen in the narrow hysteresis loop;[15,44,45] we note that a similar phase transition has been studied in Bernal bilayer and rhombohedral trilayer graphene (see Supporting Information for additional discussion).[63,64]

We theoretically model this scenario (taking $k_B = \hbar = 1$ throughout) for the case of UTe$_2$ by first considering a Ginzburg-Landau theory describing the MM phase transition:[62,65,66]

$$\mathcal{F}[\mathbf{M}](\mathbf{H}) = \frac{1}{2}\chi_i^{-1}M_i^2 + \frac{1}{4}\beta_{ij}M_i^2M_j^2 + \frac{1}{6}\gamma M_y^6 - \mathbf{M} \cdot \mathbf{H} + \\ + \kappa_j(\partial_j M_j)^2 \tag{1}$$

where $i, j = x, y, z$ that correspond to the crystallographic $a, b, c$ directions, respectively, $\mathbf{M}$ is the magnetic order parameter, while $\chi_i^{-1}, \beta_{ij}, \gamma$ and $\kappa_j$ are Ginzburg-Landau parameters. Good agreement with the experimental data is obtained only if $\beta_{xy}$ is non-zero (see caption of Fig. 7 for parameter values). We chose the parameters such that at zero applied field, the free energy has two minima: a global minimum at $\mathbf{M} = 0$, and a minimum with higher energy at $\mathbf{M} = \mathbf{M}_*$ pointing along the $b$ direction. As a magnetic field is applied, the minimum at $\mathbf{M}_*$ decreases until it becomes the new global minimum at the metamagnetic phase transition point $H_m$. We denote the energy at this minimum as $\Omega_*(\mathbf{q})$. We find that with the free energy expressed in (1), for magnetic fields aligned within the crystallographic $ab$ and $bc$ planes, a good fit is given by

$$\Omega_*(\mathbf{q}) \approx g(H_m - H_y) + \alpha H_x^2 + \sum_j \kappa_j q_j^2, \qquad (2)$$

where $g$ is a constant with dimensions of the magnetic field, and $\alpha$ is a dimensionless constant (in particular, within this approximation $\Omega_*(\mathbf{q})$ is independent of $H_z$ when $H_z \neq 0$ and $H_x = 0$). To include the effect of fluctuations on superconductivity about this minimum, we quantize the associated mode as a bosonic field $m_\mathbf{q}$, a massive magnon we refer to as a "metamagnon," with Hamiltonian $\mathcal{H}_M = \sum_\mathbf{q} \Omega_*(\mathbf{q}) m_\mathbf{q}^\dagger m_\mathbf{q}$. The metamagnon couples to the electron spin $\mathbf{S}(\mathbf{q}) = \sum_{\mathbf{k} s_1 s_2} c_{\mathbf{k}+\mathbf{q} s_1}^\dagger (\boldsymbol{\sigma})_{\alpha\beta} c_{\mathbf{k} s_2}$ (where $s_1, s_2 = \uparrow, \downarrow$ are spin indices) as $\mathcal{H}_{m,el} = \mu_e \sum_\mathbf{q} (m_\mathbf{q} + m_{-\mathbf{q}}^\dagger) S_\parallel(\mathbf{q}) M_*$, where $S_\parallel(\mathbf{q}) = \mathbf{S}(\mathbf{q}) \cdot \mathbf{M}_*/M_*$, and $\mu_e$ is the electron magnetic moment. Integrating out the metamagnon $m_\mathbf{q}$ (see Supporting Information for details) gives rise to the usual ferromagnetic spin-fluctuation interactions $\mathcal{H}_{int} = \sum_\mathbf{q} J(\mathbf{q}) S_\parallel(\mathbf{q}) S_\parallel(-\mathbf{q})$, where

$$J(q) = -\frac{\mu_e^2 M_*^2 \Omega_*(\mathbf{q})}{\Omega_*^2(\mathbf{q}) + \Gamma_m^2}.$$

Here we account for disorder via the metamagnon decay rate $\Gamma_m$ (details given in the Supporting Information). Crucially, $J(\mathbf{q}) < 0$ is an increasing function of $H_y$ and $J(0)$ is maximized at the metamagnetic phase transition.

Solving the linearized gap equation, we find that the superconducting order parameter expressed in the $\mathbf{d}$-vector notation is $\Delta(\mathbf{p}) = \mathbf{d}(\mathbf{p}) \cdot \boldsymbol{\sigma} i\sigma^y$, with $d_x = -id_z$ and $d_y = 0$ and $d_x(\mathbf{p}) = p_j$ with $j = x, y, z$ corresponding to the largest $\kappa_j$ parameter. We do not speculate which $\kappa_j$ is the largest as there are insufficient data to determine it; however, we note that possible forms of the order parameter we find include the non-unitary paired state proposed for UTe$_2$ in ref.[67] (belonging to the $B_{1u} + iB_{3u}$ irreducible representation of $D_{2h}$), as well as that considered in ref.[54] in order to explain the field direction sensitivity of the SC2 phase.

For any form of the parameter, the critical temperature for SC2 is given by

$$T_c^{(SC2)}(\mathbf{H}) = 1.13\Lambda \exp\left[-\frac{\left(\Omega_*^2(0) + \Gamma_m^2\right)^2}{8\nu\tilde{\kappa}\mu_e^2 M_*^2 \Omega_*^2(0)}\right], \quad (3)$$

where $\nu$ is the density of states, $\Lambda$ is the energy cutoff, and $\tilde{\kappa}$ is equal to the largest $\kappa_j$ times some form factor with units of momentum squared coming from integration over momentum. The corresponding $T_c$ vs $H_y$ plot is shown in Fig. 7A, which also shows a cartoon picture of $T_c^{(SC1)}$ in the SC1 phase (see Supporting Information for details). Note that in Fig. 7A we extrapolated (3) all the way up to $H_y = H_m$, though the formula is not strictly valid at that point as the coupling becomes strong.

Here we neglected several other effects that give SC2 additional dependence on the direction and strength of the magnetic field. First, fields pointing away from the $b$-axis have a component parallel to the $\mathbf{d}$-vector, and therefore suppress SC2 as does the orbital effect; we find, however, that these effects do not significantly alter the phase diagram. Second, the magnetization $\mathbf{M}_*$ of the polarized paramagnetic phase is itself a function of the applied field and changes both magnitude and direction, which in turn alters the direction of the $\mathbf{d}$-vector. Third, we have neglected any mixing between SC1 and SC2, which necessarily occurs due to the breaking of crystalline symmetries by fields aligned away from the $b$-axis. And finally, we assumed the high energy cutoff is independent of the applied field, though it is likely a function of $\Omega_*$.

In modelling the effects of disorder, we find that it is crucial that the metamagnon decay rate $\Gamma_m$ depends on the direction of the applied magnetic field, in particular if the decay is dominated by two magnon scattering and/or Gilbert damping processes.[68–70] The exact functional form depends on the precise decay mechanism, but we find phenomenologically that the data is well described with $\Gamma_m = \gamma_x \sin^4 \phi + \gamma_z \sin^4 \theta$, where $\phi$ and $\theta$ are the angles between the direction of magnetic field and the $b$−axis in the $ab$− and $bc$−crystallographic planes, respectively. The resulting phase diagram in Fig. 7B is in good qualitative agreement with the experimental data. This model therefore shows how critical magnetic fluctuations may provide the pairing glue for forming the SC2 state as the MM transition is approached – and why this phase is so sensitive to both the magnetic field tilt angle and the degree of crystalline disorder.

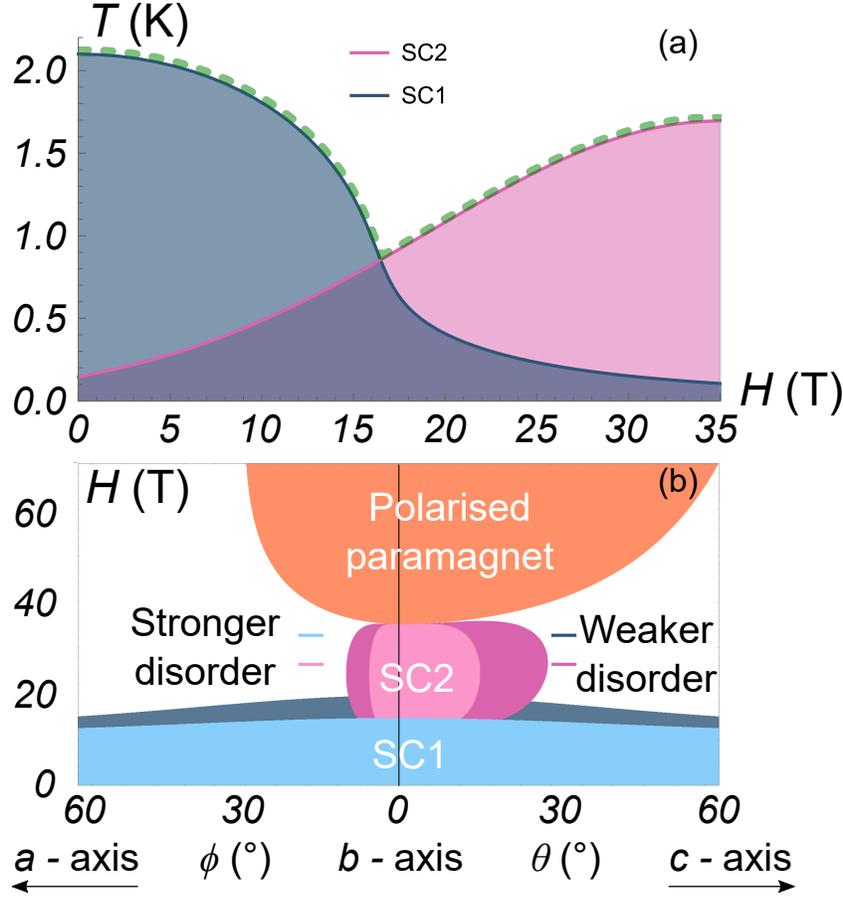

**Fig. 7.** Modelling the sensitivity to disorder of p-wave superconductivity in UTe$_2$. *(A)* Magnetic field dependence of critical temperatures for superconducting phases SC1 and SC2 for **H** oriented along the $b$–axis estimated from (3) using $\Lambda = 1.5$ K, $\frac{8M_*^2 \nu \kappa}{g^2} = 500$ and $H_m = 35$ T. Note that our linearized gap equation approach is only valid for the higher $T$ transition for all $H$ – therefore the profile of the SC1 (SC2) curve at high (low) $H$, indicating non-zero $T_c$, is outside the domain of validity of this simple modelling. The green dashed line is an envelope of the two transition lines measured experimentally, which matches well with our measurements and those reported previously.[30] *(B)* Calculated angular magnetic field phase diagram. The color coding is the same as for the experimental phase diagram in Fig. 6. The MM phase transition is obtained from Eq. (1) and is well fit with $\chi_y^{-1} = 808$, $\chi_z^{-1} = 404$, $\chi_x^{-1} = 8.08$, $\beta_{xx} = 16.16$, $\beta_{yy} = -1616$, $\beta_{zz} = 1616$, $\beta_{xy} = 16160$, and $\gamma = 646.4$ (all other parameters set to zero), with magnetic field in Tesla. We used Eq. (3) with $\Omega_*$ from Eq. (2), with resulting parameters $H_m = 35$ T, $g = 1.6 \times 10^{-3}$ T and $\alpha = 3.8 \times 10^{-5}$. For the metamagnon decay rate $\Gamma_m = \gamma_x \sin^4 \phi + \gamma_z \sin^4 \theta$, we took $\gamma_x = 0.4$ and $\gamma_z = 0.007$ to model the MSF samples and $\gamma_x = 4$ and $\gamma_z = 0.07$ to model the CVT samples (i.e. $\Gamma_m(\text{CVT}) = 10\Gamma_m(\text{MSF})$), taking $T = 0.1$ K. A good agreement between the theoretical model and experimental data is observed for both panels.

## Discussion

It is likely that a significant contributory factor to the enhancement of $T_c$ for MSF-grown UTe$_2$ is the minimization of uranium vacancies. Recent x-ray diffraction (XRD) studies on UTe$_2$ specimens of varying quality found that CVT samples with 1.5 K $\leq T_c \leq$ 2.0 K possessed uranium site defects of between $\approx$ 1-3%, while low quality samples that did not exhibit SC1 superconductivity at temperatures down to 0.45 K showed uranium vacancies of $\approx$ 4-5%.[32,40,71] By contrast, an MSF specimen with $T_c$ = 2.1 K exhibited no uranium deficiency within the experimental resolution of the XRD instrument.[32]

Therefore, the enhancement of $T_c(H)$ of the SC1 phase for field applied along each crystallographic direction, as reported for measurements of MSF samples in ref.[33] and reproduced here, is likely to be due to the minimization of uranium site vacancies for this alternative growth process utilizing a salt flux. Our striking observation of the enhanced angular profile of the SC2 phase can be well described by considering the effects of disorder on MM fluctuations, as we outlined in Fig. 7.

It has been proposed in ref.[19] that the SC2 phase may be spin-singlet in character, rather than spin-triplet as widely considered by other studies.[11,47,48,54,56–58,60,66,72,73] The authors of ref.[19] argue in favor of a singlet pairing mechanism for SC2 based on the profile of their high field specific heat measurements performed on CVT specimens. However, recent NMR measurements up to maximal applied field strengths of 32 T argue strongly in favor of SC1 and SC2 both being spin-triplet.[56–58] Interestingly, the field dependence of the $^{125}$Te-NMR intensity reported in refs.[56,57] indicates that in the SC1 phase the dominant spin component of the triplet pair points along the *a*-axis, while measurements at higher fields show that in the SC2 state the spins are instead aligned along the *b*-axis. This scenario is fully consistent with our MM fluctuation model. The broader profile of the SC2 superconducting transition (compared to that of SC1) observed in specific heat measurements in ref.[19] fits this picture of strong magnetic fluctuations near $H_m$ driving the formation of the SC2 phase. We note that superconductors in which the pairing is understood to be driven by strong nematic fluctuations also exhibit broad anomalies in their specific heat upon transitioning between their superconducting and normal states.[74,75] We propose that strong metamagnetic fluctuations underpinning SC2 thus provide a natural explanation for the observed stark difference in profile of the SC1 and SC2 specific heat transitions reported in ref.[19] However, further empirical guidance, particularly from thermodynamic probes, is evidently required in order to enable the microscopic details of the remarkable SC2 phase of UTe$_2$ to be unpicked with greater confidence and in clearer detail than we attempt here with our phenomenological model.

An interesting question posed by the observation of higher $T_c(H)$ for the SC1 phase of MSF UTe$_2$, and the purity-driven enhancement of the angular range of the SC2 phase, concerns the dependence of the SC3 state on the extent of crystalline disorder. It has recently been observed that a very low quality sample with a RRR of 7.5, which does not exhibit SC1 superconductivity down to $T = 110$ mK, nevertheless exhibits SC3 superconductivity in high magnetic fields at $T > 0.5$ K.[76] This robustness to disorder of the SC3 phase implies that it is likely very different in character to the SC2 phase, which as we have shown is highly sensitive to crystalline quality. Furthermore, whereas we found the angular extent of the SC2 phase to be considerably extended in high quality MSF samples compared to prior studies on CVT specimens, we found no evidence suggesting the angular domain, or magnetic field extent, of the SC3 phase to be markedly different. This dichotomy between the extreme sensitivity to disorder of SC2 and the remarkable robustness of SC3 calls for further careful measurements probing the differences between these two exotic superconducting phases.

Since the optimization of the MSF growth technique for high quality UTe$_2$ specimens in 2022,[32] a number of experiments on this new generation of samples have helped clarify important physical properties of this system. These include quantum oscillation measurements that reveal the Fermi surface geometry,[24,25] NMR and thermal conductivity measurements that give strikingly different results to prior CVT studies[77,78] – providing a new perspective on the possible gap symmetry – along with Kerr rotation, muon spectroscopy and specific heat measurements that also differ from prior observations and interpretations of studies on CVT specimens.[11,26,79] We are therefore hopeful that continued experimental investigation of this new generation of higher quality crystals will provide the required empirical impetus to enable more detailed theoretical models of this intriguing material to soon be attained.

In summary, we have performed a detailed comparative study of UTe$_2$ crystals grown by the molten salt flux (MSF) and chemical vapor transport (CVT) techniques. We found that the higher critical temperatures and lower residual resistivities of ultraclean MSF crystals translated into higher critical field values than prior CVT studies. Comparatively the properties of the metamagnetic (MM) transition, located at $\mu_0 H_m \approx 35$ T for $H \parallel b$, appeared the same for both types of samples. This implies that the MM transition is a robust feature of the UTe$_2$ system that is insensitive to crystalline disorder, unlike the superconductivity. Strikingly, we found that the magnetic field-reinforced superconducting state close to this MM transition (SC2) has a significantly enhanced angular range for the cleaner MSF crystals. We propose a phenomenological model – identifying the enhanced critical magnetic fluctuations close to the MM transition as a natural pairing glue for triplet Cooper pairs – which we find accurately captures our experimental observations. Our results reveal a significantly revised high magnetic field phase diagram for UTe$_2$ in the ultraclean limit, highlighting the acute sensitivity to disorder of the remarkable field-induced SC2 superconducting phase.

## Methods

UTe$_2$ single crystals were grown by the MSF technique[32] using the methodology detailed in ref.[25] Electrical transport measurements were performed using the standard four-probe technique, with current sourced along the *a* direction. Electrical contacts on single crystal samples were formed by spot-welding gold wires of 25 μm diameter onto the sample surface. Wires were then secured in place with a low temperature epoxy. All electrical transport measurements reported in this study up to maximal magnetic field strengths ≤ 14 T were performed in a Quantum Design Ltd. Physical Properties Measurement System (QD PPMS) at the University of Cambridge, down to a base temperature of 0.5 K. Electrical transport measurements up to applied magnetic field strengths of 41.5 T were obtained in a resistive magnet at the National High Magnetic Field Lab, Florida, USA, in a $^3$He cryostat with a base temperature of 0.35 K.

Skin depth measurements were performed using the proximity detector oscillator (PDO) technique.[80] This is achieved by measuring the resonant frequency, $f$, of an LC circuit connected to a coil of wire secured in close proximity to a sample, in order to achieve a high effective filling factor, $\eta$. As the magnetic field is swept, the resulting change in the resistivity, $\rho$, and magnetic susceptibility, $\chi_s$, of the sample induce a change in the inductance of the measurement coil. This in turn shifts the resonant frequency of the PDO circuit, which may be expressed as

$$\frac{\Delta f}{f} \approx -\eta \frac{\delta}{d} \left( \mu_r \frac{\Delta \rho}{\rho} + \Delta \chi_s \right), \tag{4}$$

where $d$ is the sample thickness, $\mu_r = \chi_s + 1$, and the skin depth $\delta$ may be written as $\delta = \sqrt{\frac{2\rho}{\mu_r \mu_0 \omega}}$, for excitation frequency $\omega$.[80,81] Thus, the PDO measurement technique is sensitive to changes in both the electrical resistivity and the magnetic susceptibility of the sample.

Steady (dc) field PDO measurements were performed at the National High Magnetic Field Lab, Florida, USA. One set of measurements was performed in an all-superconducting magnet utilising a dilution fridge sample space, over the temperature- and field-ranges of 20-100 mK and 0-28 T. Higher temperature, higher field measurements were obtained using a resistive magnet fitted with a $^3$He sample environment. Pulsed magnetic field PDO measurements were performed at Hochfeld-Magnetlabor Dresden, Germany, down to a base temperature of 0.6 K and up to a maximum applied field strength of 70 T.

## Data availability

The datasets supporting the findings of this study are available from the University of Cambridge Apollo Repository.[82]

## Acknowledgements

We are grateful to N.R. Cooper, I. Esterlis, H. Liu, A.B. Shick, P. Opletal, H. Sakai, Y. Haga, and A.F. Bangura for stimulating discussions. We thank T.J. Brumm, S.T. Hannahs, E.S. Choi, T.P. Murphy, T. Helm, and C. Liu for technical advice and assistance. This project was supported by the EPSRC of the UK (grant no. EP/X011992/1). A portion of this work was performed at the National High Magnetic Field Laboratory, which is supported by National Science Foundation Cooperative Agreement No. DMR-1644779 & DMR-2128556 and the State of Florida. We acknowledge support of the HLD at HZDR, a member of the European Magnetic Field Laboratory (EMFL). The EMFL also supported dual-access to facilities at MGML, Charles University, Prague, under the European Union's Horizon 2020 research and innovation programme through the ISABEL project (No. 871106). Crystal growth and characterization were performed in MGML (mgml.eu), which is supported within the program of Czech Research Infrastructures (project no. LM2023065). We acknowledge financial support by the Czech Science Foundation (GACR), project No. 22-22322S. T.I.W. and A.J.H. acknowledge support from EPSRC studentships EP/R513180/1 & EP/M506485/1. T.I.W. and A.G.E. acknowledge support from QuantEmX grants from ICAM and the Gordon and Betty Moore Foundation through Grants GBMF5305 & GBMF9616 and from the US National Science Foundation (NSF) Grant Number 2201516 under the Accelnet program of Office of International Science and Engineering (OISE). D.V.C. acknowledges financial support from the National High Magnetic Field Laboratory through a Dirac Fellowship, which is funded by the National Science Foundation (Grant No. DMR-1644779) and the State of Florida. A.G.E. acknowledges support from the Henry Royce Institute for Advanced Materials through the Equipment Access Scheme enabling access to the Advanced Materials Characterisation Suite at Cambridge, grant numbers EP/P024947/1, EP/M000524/1 & EP/R00661X/1; and from Sidney Sussex College (University of Cambridge).